\documentclass[10pt]{iopart}
\usepackage{iopams}
\usepackage{color}
\usepackage{bm} 
\usepackage{graphicx}
\usepackage[font=scriptsize, labelfont={sf,bf}, margin=1cm, justification=centering]{subcaption}
\usepackage[hidelinks]{hyperref}
\usepackage[square,numbers]{natbib}

\begin{document}

\title{Characterization of edge and scrape-off layer fluctuations using the fast 
Li-BES system on COMPASS}
\author{A. Bencze$^1$, M. Berta$^2$, A. Buz\'as$^1$, P. Hacek$^3$, J. Krbec$^3$,  M. Szuty\'anyi$^2$ and the COMPASS Team}
\address{$^1$ Wigner Research Center for Physics, Budapest, Hungary}
\address{$^2$ Szech\'enyi Istv\'an University, Gy\H or, Hungary}
\address{ $^3$ Institute of Plasma Physics of the CAS, Prague, Czech Republic}
\ead{bencze.attila@wigner.mta.hu}
\vspace{10pt}
\begin{indented}
\item[]January 2019
\end{indented}

\begin{abstract}
Recently the Lithium-Beam Emission Spectroscopy (Li-BES) system on COMPASS has 
reached its full diagnostic power in terms of routine automatic operation in any 
kind of plasma scenarios and it is normally used as a standard tool for reconstruction
of ultra fast density profiles in the edge region of COMPASS plasmas. 
The purpose of this study is to investigate the 
advantages and limitations of the COMPASS Li-BES system in characterizing plasma electron density fluctuations. We show how the atomic physics of plasma-beam interactions can affect the interpretation of the measurement at different radial positions and for different electron density profiles. We also demonstrate the usability of generalized sequential probability ratio test for automatic event detection. Using non-perturbative diagnostic, we verify the validity of the stochastic Garcia-model for scrape-off layer filaments and accompanying holes (density deficits).
\end{abstract}

%
%
%
%
\ioptwocol

\section{Introduction}
Tokamak plasma region defined as a narrow, few cm band around the last-closed-flux-surface (LCFS) is considered to be of crucial importance for understanding different transport phenomena determining both particle and energy confinement times and the strength of the plasma-wall interaction \cite{ref.CARRERAS_jnm2005,ref.NAULIN_jnm2007,ref.krasheninnikov_jpp2008,ref.garcia_pfr2009,ref.dIppolito_pop2011}. 
This is the region where  the important and interesting physics of the L-H transition, the generation of turbulence induced mesoscopic shear flows takes place and where the scrape-off layer (SOL) filaments, known as 'blobs' are born. This region is well known to be rich in different stable and unstable modes like drift  waves (DW), neoclassical tearing modes (NTMs), edge localized modes (ELMs) etc. The unstable modes through nonlinear mode couplings can and will lead to coherent structure formation.  The kinematics and dynamics of these structures (which usually take the form of 3D filaments) is determined by the actual plasma parameters both global and local, the underlying shear flows and the interaction of filaments with the material components such as divertor and limiter structures.  Theoretical descriptions of the coherent structure generation, structure propagation and detection statistics have reached different levels of sophistication and descriptive power. Some of the theories are competitive \cite{ref.carreras_pop2001,ref.sanchez_prl2003,ref.carralero_ppcf2011,ref.devynck_pop2005}, therefore any experimental insight is of significant value.

The present work is intended to be a comprehensive contribution to the growing body of experimental data concerning the characterization of coherent structures: filaments or blobs in the scrape-off layer, holes or density deficits and other fluctuating phenomena in the plasma edge region. At the same time our intention is to place the recently built COMPASS Lithium Beam Emission Spectroscopy (Li-BES) system, on the map of plasma turbulence studies, demonstrating its capabilities and uncover some of the interpretation issues that may arise during fluctuation data analysis.

The rest of this paper is organized as follows: in \autoref{sec:li_beam_emission_spectroscopy}, the experimental setup and the principles of Li-BES diagnostic are discussed. \autoref{data_characterization}, qualitatively describes the raw data identifying the main noise sources. The effect of the plasma-beam interaction on the measured signal is described in details in \autoref{sec.plasma_beam_interaction}. Subsection \ref{sec:fluctuation_amplitudes_and_correlations} is devoted to the fluctuation amplitude profiles and the variation of the auto-correlation functions with the radial position. Section \ref{sec:sol_plasma_fluctuations} summarizes the analytic description of the Garcia-model (\autoref{sec:garcia}) and presents the results of the synthetic diagnostic (\autoref{sec:blob_simulation}). Before analyzing the actual experimental data, two event detection techniques are compared (\autoref{sec:blob_detection}). After presenting the experimental results in \autoref{sec:experiment}, a summary is given.  
    
\section{Li-Beam Emission Spectroscopy}\label{sec:li_beam_emission_spectroscopy}
Neutral alkali beams are routinely used in a number of fusion-related devices as non perturbative diagnostics for time evolving density profile measurements \cite{ref.fiedler_jnm1999,ref.zoletnik_rsi2018_02,ref.refy_rsi2018,ref.anda_rsi2018}.

\begin{figure}
    \includegraphics[width=83mm]{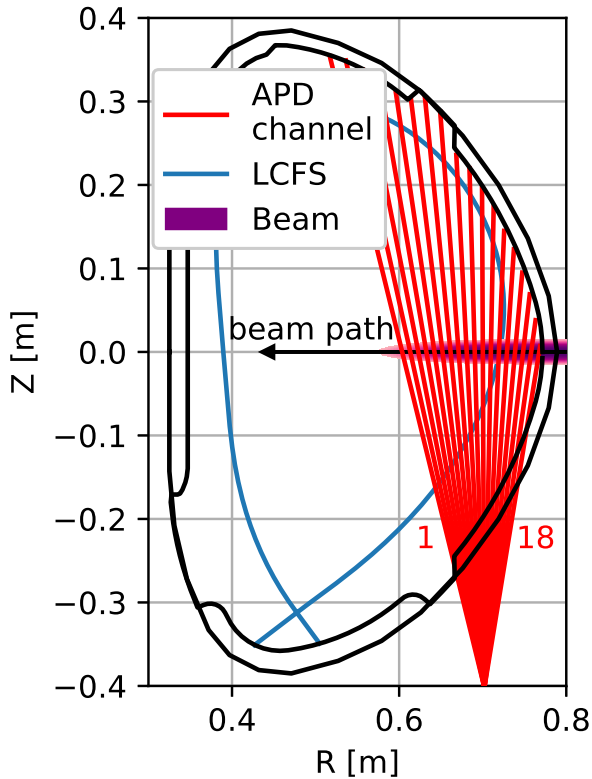}
    \caption{Experimental setup showing the Li-beam path and the observation directions (APD line of sights) relative to the tokamak vessel and the last closed surface (LCFS).}
    \label{fig:Li_beam_vessel}
\end{figure}

Experimental data analyzed in the present article have been acquired by the Li-BES diagnostic system (see Figure \ref{fig:Li_beam_vessel}) which consists of two main parts: the high energy ($60~\mathrm{keV}$) diagnostic Li-beam \cite{ref.anda_fed2016} and the Li-light detection system \cite{ref.berta_fed2015}. The Li-gun ion source emits $\mathrm{Li}^+$ ions which are further accelerated in two stages (extraction and acceleration), in our case up to $60~\mathrm{keV}$. The ion beam is focused in the ion optics before entering the neutralization chamber where it gets neutralized by charge exchange processes induced by collisions with sodium atoms.  The neutral beam reaches the plasma without any significant energy loss. During the plasma-beam interaction, through collisions with the plasma particles, Li-atoms are excited mainly to  $2\mathrm{p}$ state by the plasma particles. The excited $2\mathrm{p}$ state decays with a $\tau_\mathrm{life}\approx 27~\mathrm{ns}$ lifetime with the emission of a photon of characteristic wavelength ($\lambda=670.8~\mathrm{nm}$). These photons can be observed using an appropriate $2~\mathrm{nm}$ wide interference filter and various detector systems (CCD camera, photomultiplier, photodiode or Avalanche photodiode (APD)).

For the present analysis we use the data acquired by a 18 channel APD-detector array  \cite{ref.dunai_rsi2010} with observation volumes aligned along the major radius of the tokamak at the outer midplane. The distance of neighboring observation volumes is $10~\mathrm{mm}$. The actual radial resolution of the measurement strongly depends on the nonlocal nature of the light emission due to the finite lifetime of the exited Li(2p) states. The upper limit for this non-locality can be easily calculated as:
\begin{equation}\label{eq.delta_r}
\Delta r = \sqrt[]{\frac{2E_\mathrm{beam}\tau_\mathrm{life}^2}{m_\mathrm{Li}}},
\end{equation}
where $E_\mathrm{beam}$ is the beam energy (most of our measurements were done using $60~\mathrm{keV}$ beam energy), $\tau_\mathrm{life}$ is the lifetime of the excited Li(2p) atomic state and $m_\mathrm{Li}$ is the mass of individual Li atoms. According to the Eq. (\ref{eq.delta_r}), in our measurement $\Delta r\approx 35~\mathrm{mm}$. This value sets a theoretical maximum for the smearing effect of the beam. The effective value of this non-locality is lower due to the collision induced de-excitation and ionization  depending on the actual plasma density profiles \cite{ref.asztalos_eps2017}. In \cite{ref.willensdorfer_ppcf2014} a basic sensitivity study has been performed concerning the response of the light emission profile to the local perturbations of the density profile, concluding that the sensitivity is strongly suppressed at the maximum of the emission profile. Later in this paper we are going to analyze the effect of atomic physics on the statistical properties of fluctuating Li-BES signals.   

The detected light intensity is –- in the first approximation -– proportional to the local electron density, therefore observing the time average intensity and its fluctuations along the beam, it is possible to reconstruct the density profile and the quasi-2D correlation functions of the electron density fluctuation \cite{ref.fonck_rsi1990,ref.zoletnik_ppcf1998}.
Using a pair of deflection plates the beam can be either chopped or poloidally deflected (swept). Chopping the beam makes possible to correct for the background, while the poloidal deflection allows quasi-2D measurements \cite{ref.zoletnik_ppcf2012}. The switching frequency of the sweeping/deflection system presently works up to  250 kHz. The poloidal resolution is limited by the $\approx 2~\mathrm{cm}$ beam diameter. In the experimental campaign of which data are analyzed in this article, the quasi-2D measurement was not available. 

 \subsection{Data characterization}\label{data_characterization}
 
\begin{figure}
\includegraphics[width = 83mm]{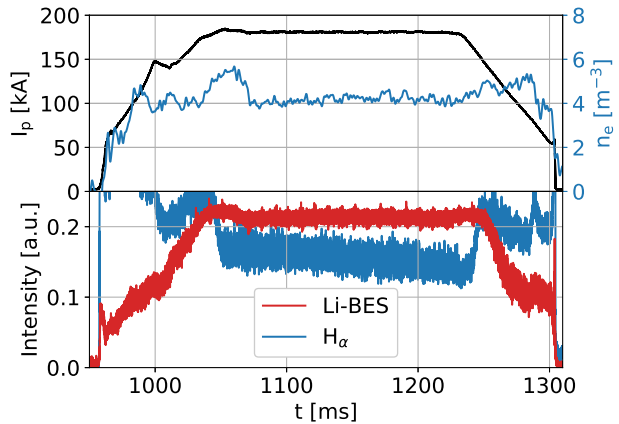}
\caption{Time traces of the main parameters of the discharge \# 16072. In addition the Li-BES time signal is shown for the brightest APD channel.}
\label{fig:16072_main_param}
\end{figure}

  Data analyzed in this work have been obtained in the experimental campaign CC18.18 
  dedicated to acquiring large sets of fluctuation data in identical discharges  in order to provide good statistics for in-depth plasma fluctuation studies. Medium density, L-mode plasmas were produced with a stable $150~\mathrm{ms}$ flat-top. The main plasma parameters of the discharges are shown in Figure \ref{fig:16072_main_param}.

    For our practical purpose we separate the outer region of the toroidal plasma in two parts: the edge plasma and the SOL.
    In this article we define the edge plasma as the region of closed magnetic surfaces just few cm inside the last closed flux surface (LCFS), meaning two observation channels ($R_{12}=709.7~\mathrm{mm}$, $R_{13}=719.4~\mathrm{mm}$) in the Li-BES system of COMPASS, while the SOL plasma is the region of open field lines and in this article we consider three Li-BES channels ($R_{15}=738.7~\mathrm{mm}$, $R_{16}=748.3~\mathrm{mm}$, $R_{17}=758.0~\mathrm{mm}$) for our SOL analysis.
    
    It is quite obvious that the quality of the 
    Li-BES data is highly correlated to the reliability of the outcome results, therefore the assessment of data quality is very important and often neglected issue in such data analysis.
    
    In experimental physics the analyzed time traces are usually thought to consist of a series of instances taken from some statistical ensemble. The assumption is that the statistics is determined by the -- usually nonlinear -- dynamics of the physical system. In real world various noise sources can corrupt and hide the interesting statistics of the system under investigation. This is very much the case with the fluctuation data detected by the Li-BES system. In our case the interesting phenomena is the fluctuating turbulent plasma flow driven by local micro instabilities, while the main sources of measurement uncertainty can be identified as:
    \begin{enumerate}
        \item Photon statistical noise
        \item Amplifier noise
        \item Pick-up oscillatory noise
        \item Fluctuations in the plasma background
    \end{enumerate}
\begin{figure}
    \includegraphics[width=83mm]{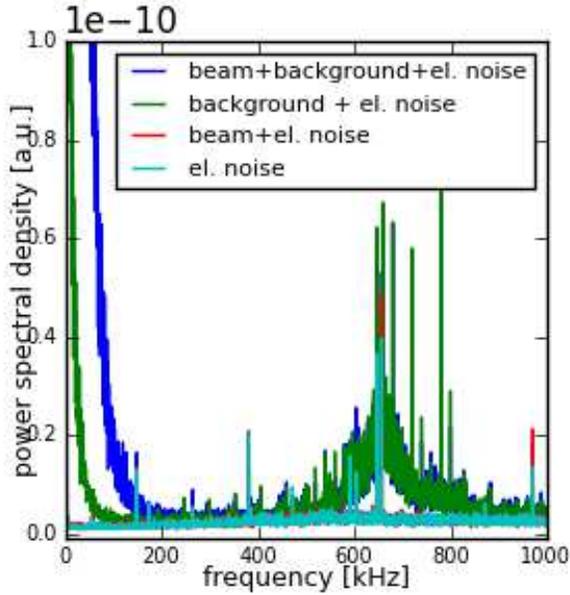}
    \caption{Power spectra of the fluctuating light signal from a SOL APD channel.}
    \label{fig:noise_spectrum}
\end{figure}
Figure \ref{fig:noise_spectrum} shows spectra of a typical Li-BES signal. The beam has been switched on and off with $20~\mathrm{ms}$ period in order to asses the effect of plasma background and of the electronic noise. After the plasma discharge ends (after shot period, ASP) the beam observation continues. When the beam is off in the ASP, the pure electrical noise is detected (see the cyan line in Figure \ref{fig:noise_spectrum}, this is also the sum of items (i)+(ii)+(iii) in the list above), some sporadic narrow peaks are seen in the spectrum. The same peaks are seen when the beam is on in ASP (see the red line in Figure \ref{fig:noise_spectrum}), showing that the presence of the beam does not introduce additional noise to the spectra. We should mention the resonance around $650~\mathrm{kHz}$ which gets amplified and broadened when the high photon flux hits the detector during the discharge (see the green and blue lines in Figure \ref{fig:noise_spectrum}). These high frequency features are irrelevant from the point of view of edge turbulence studies, since we are interested in the $1- 100~\mathrm{kHz}$ spectral range. It can be also seen, that the background fluctuation levels are at least one order of magnitude lower then the beam light fluctuations in the relevant spectral range, assuring that the detected fluctuations are well localized at the beam position.

\subsection{Effects of plasma-beam interaction}\label{sec.plasma_beam_interaction}

 As the BES system detects the light emitted by plasma-beam interactions,  the information about plasma density is indirect. A natural question arises: what is the effect of  plasma-beam interactions on statistical quantities such as fluctuation amplitude  profiles, correlation functions, conditionally averaged wave forms etc. To study the question a new reconstruction code has been developed at COMPASS \cite{ref.krbec_rsi2018} based on Schweinzer's density profile reconstruction algorithm \cite{ref.schweinzer_ppcf1992}. The method gives an approximate solution of the collisional-radiative model including electron-impact excitation, ionization and charge exchange processes:
\begin{equation}
\label{eq:CollRadModel}
\frac{\mathrm{d}}{\mathrm{d}z}N_i(z) = \sum_{j=1}^5\left[n_\mathrm{e}(z)a_{ij}(T_\mathrm{e}(z)) + b_{ij}\right]N_j(z),
\end{equation}
where $N_i$ is the occupation density of the $i^\mathrm{th}$ excitation level, $z$ is the coordinate along the lithium beam, where $z = 0$ defines the position of the first measurement channel (called channel \# 18). The coefficients $a_{ij}$ $(i \neq j)$ describe the transition rate from level $i \longrightarrow j$ due to collisions with plasma electrons and ions. 
Attenuation of the $i^\mathrm{th}$ atomic state due to charge exchange and ionization
is included in coefficient $a_{ii}$. The coefficients $b_{ij}$ are the Einstein coefficients of spontaneous emission. $n_\mathrm{e}$ and $T_\mathrm{e}$ are the electron density and temperature, respectively. This linear differential equation system describes the light emission of a Li-beam passing through a given plasma density and temperature profile.
\begin{figure}
\centering
    \includegraphics[width=83mm]{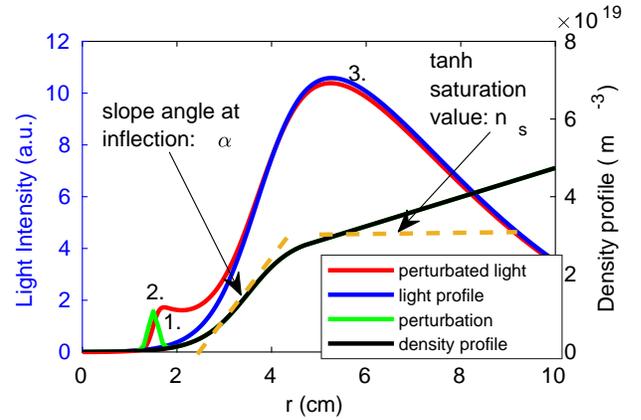}
    \caption{Artificial density profile and calculated light profile with added perturbation and response. See detailed explanation in the text.}
    \label{fig:profiles}
\end{figure}

 \begin{figure*}
\centering
    \includegraphics[width=166mm]{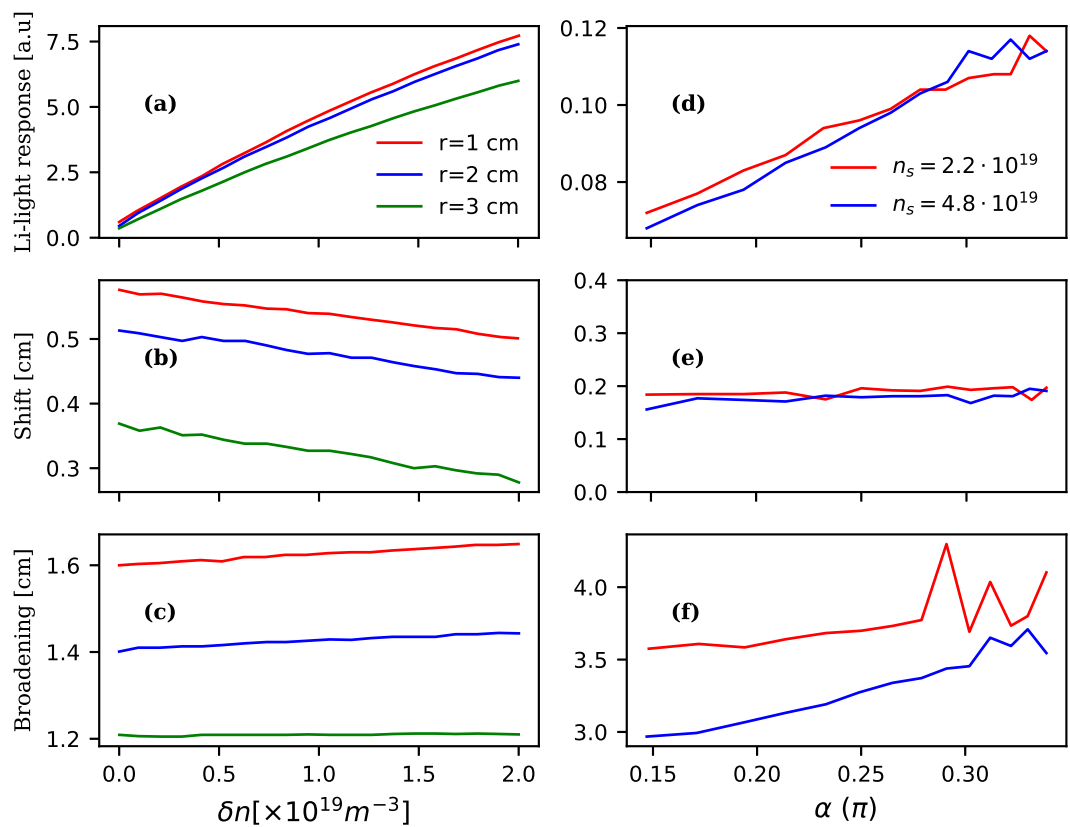}
    \caption{The amplitude response (a), the offset of the response (b) and the widening of the response (c) as a function of perturbation amplitude at different radial locations with respect to the boundary of simulation domain. The amplitude response (d), the offset of the response (e) and the widening of the response (f) as a function of profile gradient at different saturation densities (see \autoref{fig:profiles}).}
    \label{fig:pert_resp}
\end{figure*}

 \autoref{fig:profiles} shows an artificially constructed density profile (black curve) which consists of a $\tanh$ function smoothly merged to a linear function, matching the observed shapes of real density profiles \cite{ref.stefankova_rsi2016}.  The shape of the profile is controlled by two parameters: (i) the gradient at the inflection point ($\alpha$ is the angle associated with the tagent line),  (ii) the density value where the $\tanh$ function saturates ($n_\mathrm{s}$). The density profiles constructed in such way have been perturbed with localized Gaussian density pulses. Using the collisional-radiative model (\ref{eq:CollRadModel}) the light emission can be calculated. The calculations shows some distinct features of the light emission response (see \autoref{fig:profiles}):
 \begin{itemize}
     \item amplification/attenuation of the perturbation amplitude
     \item radial shift of maximum perturbed emission
     \item radial broadening of the perturbation
     \item decrease in the emission behind the light profile maximum
 \end{itemize}
 
 Recently it has been demonstrated that the relative emission response to density perturbation decreases with the radial position of the perturbation \cite{ref.willensdorfer_ppcf2014}. The present work expands on this idea as we examine the characteristics of the light response to various density perturbations of different density profiles.
 
Let us first examine  Figure \ref{fig:pert_resp}.a-c where the properties of light emission response are plotted against the perturbation amplitude. The subplot (a) reproduces the findings in \cite{ref.willensdorfer_ppcf2014}: the amplitude response is linear for small perturbations and a small deviation is observed for larger $\delta n$. It is also clear that for a given perturbation strength the response decreases as we travel uphill the profile. According to Figure \ref{fig:pert_resp}.b  the maximum of the perturbation is shifted towards the center by a considerable amount of $\approx 0.5~\mathrm{cm}$ for low perturbation amplitudes. Less radial shift is observed for larger amplitudes and for larger $r$ values. Depending on the amplitude and the initial position of the perturbation, a well localized perturbation gets broadened (smeared) due to the finite lifetime of the exited Li(2p) atomic states. 
Figure \ref{fig:pert_resp}.c reveals that larger $r$ values favour localization. 

Figure \ref{fig:pert_resp}.d-f presents the situation when a fixed amplitude density
perturbation is applied to the density profile at the radial
position where the $\tanh$ profile reaches its inflection point (maximum gradient). The response has been calculated for two $n_s$ saturation densities and different profile gradients $\alpha$ (see Figure \ref{fig:profiles}).
As the density profile becomes steeper the sensitivity to a given perturbation increases independently of the saturation density (see Figure \ref{fig:pert_resp}.d), while the radial shift of the perturbation does not depend neither on $\alpha$ nor on $n_s$ as shown in Figure \ref{fig:pert_resp}.e. For low densities the perturbation broadening is quite large and it is basically independent of the profile gradient, on the other hand for higher density and shallow profiles the smearing is considerably smaller due to the collisional deexcitation processes. 
 
\subsection{Fluctuation amplitudes and correlations}\label{sec:fluctuation_amplitudes_and_correlations}

\begin{figure}
\begin{center}
\includegraphics[width=83mm]{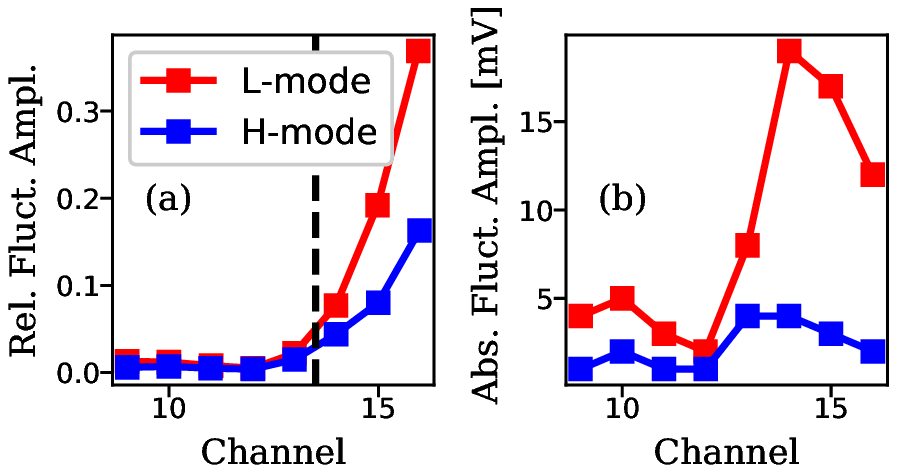}\\
\includegraphics[width = 83mm]{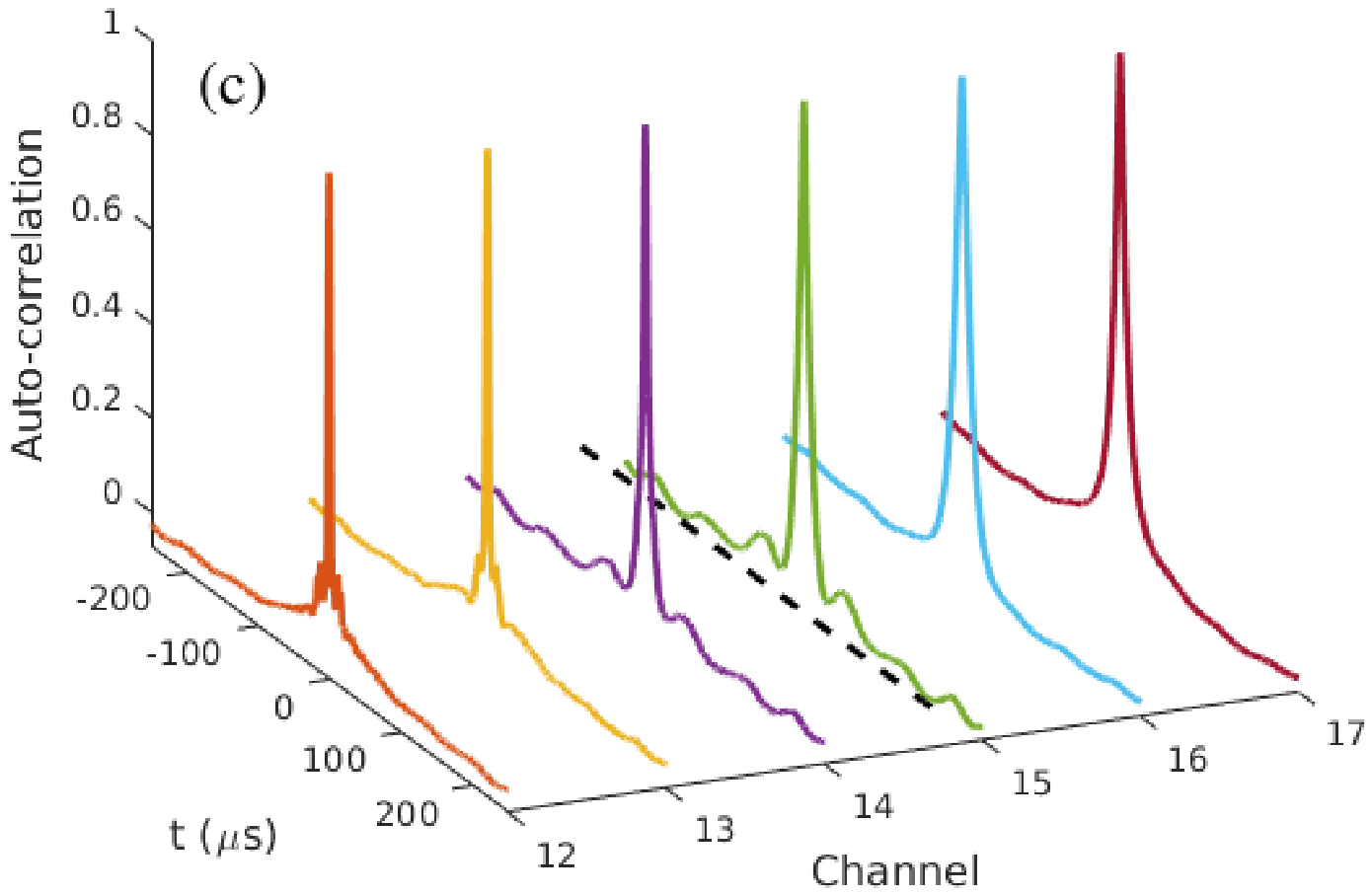}
\caption{Relative (a) and absolute (b) fluctuation amplitude profile of Li-light signal in L (red) and inter-ELM H-mode (blue). Radial variation of the auto-correlation function shape (c). The dashed black line depicts the separatrix position.}
\label{fig:fluct_light}
\end{center}
\end{figure}

Before we turn to the detailed analysis of scrape-off layer structures (density filaments/blobs, holes), we shall describe the global characteristics of measured Li-light fluctuations (as a proxy for the electron density fluctuations).
The relative and absolute fluctuation levels are plotted in \autoref{fig:fluct_light}.a-b respectively, and are calculated as:  

\begin{equation}
\delta \mathrm{I}_{\mathrm{abs}}=\sigma_a - \sigma_b,
\end{equation}

\begin{equation}
\delta \mathrm{I}_{\mathrm{rel}}=\frac{\delta \mathrm{I}_{\mathrm{abs}}}{\mu_a - \mu_b},
\end{equation}
where $\sigma_a$ and $\mu_a$ are respectively the standard deviation and the time average of the signal $S_a (t) = S_\mathrm{beam}(t)+S_\mathrm{background}(t)$. This is the superposition of plasma background emission and the Li-beam emission. $\sigma_b$ and $\mu_b$ are the standard deviation and the time average of plasma background signal $S_b(t) = S_\mathrm{background}(t)$ respectively. Applying a properly designed FIR bandpass filter, $S_a(t)$ and $S_b(t)$ contain only the relevant frequency components in the range of $f\in [1,100]~\mathrm{kHz}$.

A clear difference between L-mode and H-mode fluctuation strength is visible in both absolute and relative fluctuation amplitude. The absolute fluctuation level in L-mode is as much as five times higher than in H-mode, while the relative fluctuation amplitude shows a factor of two difference. This behaviour is well known and generally observed in all toroidal plasma devices (see e.g \cite{ref.zweben_ppcf2007} and the references therein). The lower plot in \autoref{fig:fluct_light} shows the radial variation of auto-correlation function shapes. In the scrape-off layer fluctuations with $10\% - 40\%$ relative level are seen with auto-correlation functions smoothly decaying with a decay length $\tau_\mathrm{decay}\in[20,50\mu\mathrm{s}]$ decreasing towards the separatrix. In the last closed flux surface region (plasma edge) the fluctuations are different as can be seen in \autoref{fig:fluct_light}.c. The observed  wave-like fluctuations with $\approx 70~\mathrm{kHz}$ frequency are similar to the earlier Li-BES observations at Wendelstein 7-AS stellarator \cite{ref.zoletnik:pop1999}. 

\section{SOL plasma fluctuations}\label{sec:sol_plasma_fluctuations}
In this section we give a detailed description of the observed statistics of scrape-off layer fluctuations measured by Li-BES. Prior to the discussion of the blob detection and statistical analysis,  we present the stochastic modeling of Garcia and a computer simulation based on this model,  simulating both the density and Li-light emission fluctuations.  

\subsection{The Garcia model}
\label{sec:garcia}

In \cite{garcia2012} Garcia presented a stochastic model  describing the fluctuations of scrape-off layer plasma as perceived by single point measurements. The observed behavior is dominated by large-amplitude bursts, commonly referred to as "blobs". The model represents the measurement as a random sequence of bursts:
\begin{equation}
 \Phi(t) = \sum_kA_k\psi(t-t_k),
   \label{eq:shot_noise}
 \end{equation}
where $A_k$ is the burst amplitude, $t_k$ is it's arrival time and $\psi(t)$ is a fixed waveform. It also states that bursts arrive according to a Poisson-process, which 
means that the waiting time between them follows an exponential distribution with rate $\frac{1}{\tau_w}$:
\begin{equation}
 P_{\tau}(\tau) = \frac{1}{\tau_w}e^{-\frac{\tau}{\tau_w}}.
   \label{eq:Poisson}
 \end{equation}
Calculating the cumulants of the signals probability density function (or PDF: $P_{\Phi}(\Phi)$) yields formulas for its skewness and flatness, that show a parabolic relation, 
which is characteristic of systems that are dominated by intermittent fluctuations:
 \begin{equation}
  F = 3 + \frac{I_2I_4}{I_3^2}\frac{\langle A^2\rangle\langle A^4\rangle}{\langle A^3\rangle^2}S^2,
   \label{eq:skew}
 \end{equation}
where $I_n$ is the integral of the $n^{th}$ power of the wave form:
 \begin{equation}
  I_n = \int_{-\infty}^{\infty}\mathrm{d}t[\psi(t)]^n 
   \label{eq:In}
 \end{equation}
The possible origin of such quadratic relation has been discussed in \cite{ref.guszejonov_pop2013}. For experimental consideration the burst waveform is taken as a "step" rise followed by an exponential decay: $\psi(t) = \Theta(t)e^{-\frac{t}{\tau_d}}$ and exponentially 
distributed burst amplitudes. The connection between the skewness and flatness thus simplifies to:
 \begin{equation}
F = 3 + \frac{3}{2}S^2.
   \label{eq:skew_simple}
 \end{equation}
In case of short waiting times and slow decay ($\gamma = \frac{\tau_d}{\tau_w}$ is large) $P_{\Phi}(\Phi)$ approaches a normal distribution while in the opposite limit it approaches 
Gamma distribution:
 \begin{equation}
\lim_{\gamma\rightarrow0}P_{\Phi}(\Phi)=\lim_{\gamma\rightarrow0}\frac{1}{\Gamma(\gamma)}\frac{1}{\Phi}e^{-\frac{\Phi}{\langle A\rangle}}.
   \label{eq:gamma}
 \end{equation}
In \cite{garcia2016} a slightly refined version of the model was presented according to experimental results from Langmuir-probe measurements on the TCV tokamak. The shape of the bursts that dominate the signal have a double exponential waveform with amplitude and waiting time both fitting exponential distributions. The PDF of the signal 
fits a Gamma distribution which is in agreement with previous results \cite{garcia2012}.

\subsection{Simulation and synthetic diagnostic}\label{sec:blob_simulation}

We created a MATLAB  simulation of intermittent SOL fluctuations based on the Garcia model described in \autoref{sec:garcia}. This simulation connected to the atomic physics model described previously serves as a synthetic diagnostic helping the interpretation of the Li-BES measurements.  
It basically models a single point measurement in a two-dimensional (radial-poloidal) plane. 
The basic parameters of the emerging blobs are randomly generated: the arrival times at the first radial observation point and the blob amplitudes are drawn from an exponential probability distribution. Individual blobs are initialized at fixed radial position $r_0$ ($r=0$ is the outer boundary of the simulation domain) and moved with velocities linearly dependent on blob amplitude, scattered by a normal distribution to simulate the findings in \cite{garcia2016}.
%

Blob wave fronts are propagated radially outward while their amplitudes are exponentially decaying. For each measurement point the  \emph{passing through} time (blob duration) is calculated for every blob, than an exponential rise and fall is added to the signal, creating double exponential waveforms. 
%

\begin{figure}
\centering
    \includegraphics[width=83mm]{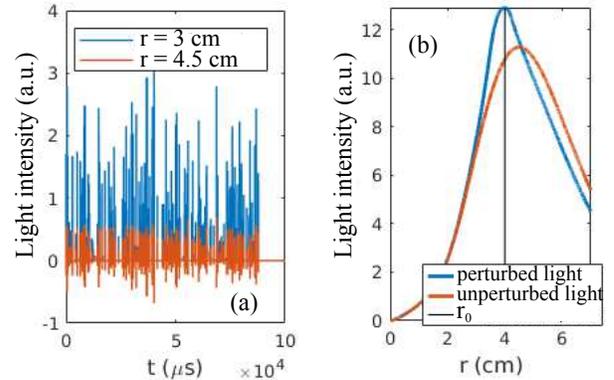}
\caption{Time traces of simulated Li-emission signals at two different radial positions  (a). Effect of atomic physics on a single blob around $r_0$ (b).}
\label{fig:sim_light_fluc}
\end{figure}

\begin{figure}
\centering
    \includegraphics[width=83mm]{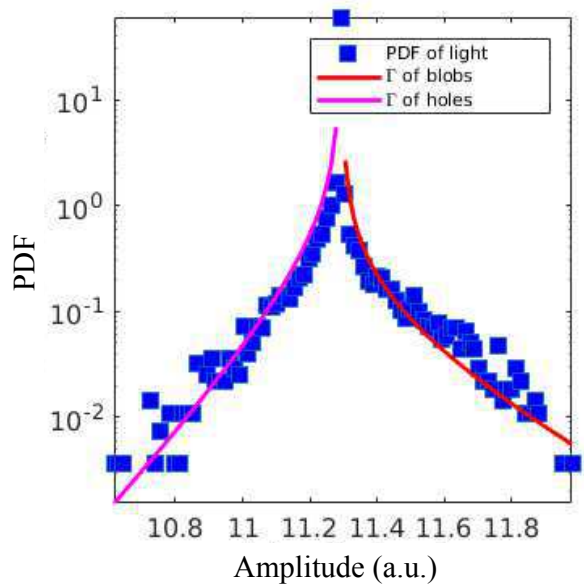}
\caption{The PDF of a simulated signal, where both blobs and artifact holes are present ($r>r_0$). The red and magenta lines are fitted Gamma distribution according to the predictions in \cite{garcia2012}.}
\label{fig:hole_pdf}
\end{figure}

The poloidal shape of individual blobs is described by a Gaussian function with $\mathrm{FWHM} = 3~\mathrm{cm}$. Then the perturbed signal is added to a radial density profile reconstructed from Li-BES measurements \cite{ref.krbec_rsi2018}. The amplitudes are scaled to match the relative fluctuation amplitudes seen in the experiment.
%
The average blob detection frequency is set 
so that larger blobs arrive at around $1~\mathrm{kHz}$. The shape parameters and the decay of the blobs is ascertained using conditionally averaged waveforms (conditional averaging will be explained in the next subsection) from COMPASS experiments as well. The radial velocity of the blobs is set to be around $300~\frac{m}{s}$ as in e.g. \cite{ref.lampert_pop2018}. The average blob amplitude is arbitrary as it can be scaled while maintaining statistics. 
After simulating a fluctuating density signal, the atomic physics of plasma-beam interaction is taken into account as it has been described in subsection \ref{sec.plasma_beam_interaction}. This way we get a synthetic signal similar to what we can obtain from real Li-BES measurements (see \autoref{fig:sim_light_fluc}). 
At first glance two important features can be noted: (i) negative spikes appear at $r>r_0$, where $r_0$ is the radial coordinate, where all blobs are created (in this simulation $r_0=4~\mathrm{cm}$). The negative spikes or holes are artifacts caused by the beam loss due to increased ionization in the higher density zone.
(ii) the positive spikes at $r>r_0$, but closer to $r_0$ are the result of the smearing of the density perturbation by the finite lifetime of the excited Li(2p) states.
In Figure \ref{fig:sim_light_fluc}, the subplot (a) depicts two time traces at different radial positions: the blue curve displays only positive fluctuations at $r<r_0$, while the red curve shows both positive and negative fluctuations relative to the average light profile. It is very important to realize that the appearance of holes as artifacts can corrupt our interpretation of the experimental results, therefore extra care is required during BES data analysis. Their contribution to the PDF (Probability Density Function approximated by the amplitude histograms) of the signal also fits a Gamma distribution predicted by \cite{garcia2012} (as seen on \autoref{fig:hole_pdf}).

\begin{figure}
\centering
    \includegraphics[width=83mm]{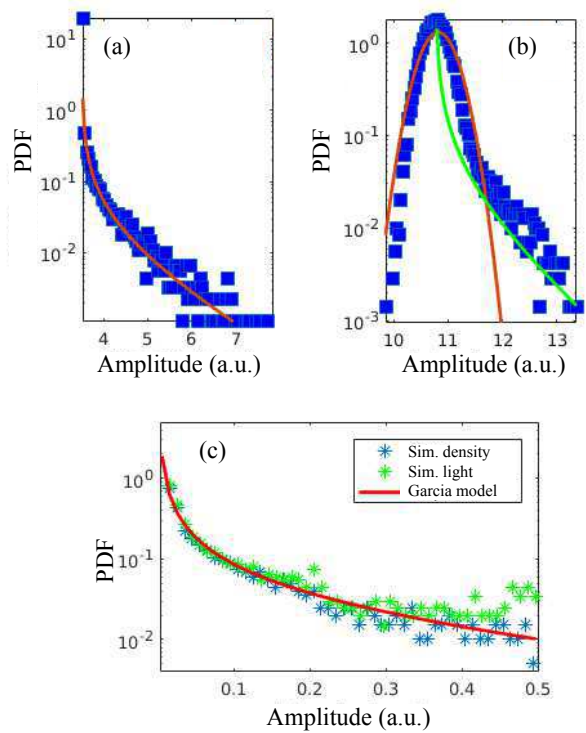}
\caption{(a) PDF of simulated density fluctuations (red line: fitted Gamma function), (b) PDF of light fluctuations with Gaussian noise (red line: fitted normal distribution, green line: fitted Gamma distribution), (c) comparison of the density fluctuation PDF  with the Li-light fluctuation PDF (red line:  fitted Garcia model prediction).}
\label{fig:sim_dens_light_pdf}
\end{figure}

 We applied a broadband Gaussian random noise to the simulated "density" signal with standard deviation proportional to the signal RMS. This accounts for the different noises captured by the observation system, such as the electronic noise and the photon statistical noise in the limit of high photon fluxes. As a reference, in Figure \ref{fig:sim_dens_light_pdf}.a  the PDF of the "pure" density signal is shown. A Gamma distribution fits perfectly the simulation. 
The resulting PDF of the noisy synthetic measurement (Li-light fluctuations) is shown in Figure \ref{fig:sim_dens_light_pdf}.b, and exhibit a PDF with Gaussian center (red curve) and  Gamma-like tail (green curve). We can conclude that  applying the effect of collisional-radiative processes to the density does not change the amplitude distribution significantly, the change in PDF shape (observed in the experiments) is introduced by the detection noise. See \autoref{fig:sim_dens_light_pdf}.c, where  scaled amplitude distributions are showed for the simulated noise-free density and light fluctuations. The Gamma distribution predicted by the Garcia model accurately fits the simulation density as well as Li-light data. 

\subsection{Event detection: two approaches}\label{sec:blob_detection}

The most frequently used method in the literature for SOL blob detection is based on a predetermined, otherwise arbitrary threshold. This threshold is usually given in terms of the standard deviation $\sigma$ \cite{garcia2016}. The detection algorithm scans the signal and finds the peaks above some prescribed  $\sigma$ value. The detected peaks are considered to be part of individual filamentary events, therefore we can calculate the detection frequency, the amplitude distribution and the waiting time distribution. This approach is used in this article for conditional averaging (see section \ref{sec:conditonal_average}).

Another recently introduced approach to detect Edge Localized Mode (ELMs) events in different diagnostic signals is the generalized Sequential Probability Ratio Test (gSPRT) \cite{ref.berta_fed2017}. In the present work we applied, for the first time, the gSPRT technique to blob detection. It is a unique characteristic of the method that the event duration can be determined accurately according to the intrinsic statistics of the experimental data.

\subsubsection{The gSPRT method.}
The generalized method described in \cite{ref.berta_fed2017} is based on the classical Sequential Probability Ratio Test \cite{ref.wald_AnnMathStat1945}, but adapt the local physics conditions and get the required parameters from the measured diagnostic signal.
The SPRT algorithm is based on the assumption that the signal consists of two separate random processes and the algorithm can decide for each sample what is the logarithmic probability ratio of the two situations where the sample belongs to one or the other processes. This parameter, usually denoted by $\lambda_n$ (see its definition in \cite{ref.berta_fed2017}), depends on the probability distributions (PDF) of the two processes. Two decision levels are defined for $\lambda_n$, discriminating between two processes:
\begin{eqnarray}
A &=& \ln\frac{\beta}{1-\alpha},\\
B &=& \ln\frac{1-\beta}{\alpha},
\end{eqnarray}
where we set  $\alpha$ - the probability of 'false alarm' – to $0.2$ and $\beta$ – the probability of 'missed alarm' to $0.05$. We have to note that the gSPRT method is not sensitive to small variations of these values. When $\lambda_n$ is below level $A$ the sample belongs to the first process (here inter-blob fluctuations), while it is above level $B$ the sample belongs to the second process (here the blob event).

\begin{figure}
\centering
    \includegraphics[width=83mm]{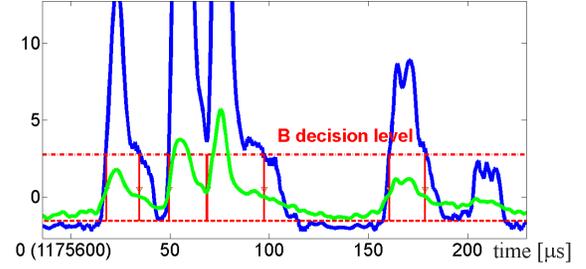}
    \caption{The calculated gSPRT signal (blue) together with the original signal (green) showing the starting and ending points as calculated by the gSPRT algorithm.}
    \label{fig:gSPRT}
\end{figure}

For defining $\lambda_n$ directly from the experimental data we use the best fit two-parameter amplitude distribution (ADF) as a proxy for the PDF. We have found that the ADF of the inter-blob periods can be well described by the Gaussian distribution $G(\mu_{0}$, $\sigma_{0})$. On the other hand, the best fit for the empirical ADF for conditionally averaged “blob event” was the Inverse Gauss (also known as Wald) distribution $W(\mu, \Lambda)$.
We have found a strong correlation ($>95\%$) between the maxima of the calculated gSPRT signal ($\lambda_n$) for each blob and the maxima of the original signal. As it is shown in Fig. \ref{fig:gSPRT} the gSPRT can determine the start and the end of the 'blob event'. 

When the gSPRT signal ($\lambda_n$) exceeds the decision level $B$, the blob event starts and when the signal falls below the level the event ends. Figure \ref{fig:gSPRT} also illustrates the problem of merged events. When two or more events overlap, the gSPRT signal could not fall below the decision level $B$ and the algorithm could not distinguish between the events. This problem can be handled by finding the local minima of the smoothed gSPRT signal within a given merged blob, and put the end and the start points there as it can be seen in the case of the 2nd and 3rd blob in the Figure \ref{fig:gSPRT}. We label the detected blobs according to their intensity (size). E.g. a 'size 2' blob has its maximal amplitude in the corresponding gSPRT signal in the interval $[B+1\cdot\Delta l,B+2\cdot\Delta l]$, where $\Delta l = (M-B)/n$ and $M$ is the overall maxima for the gSPRT signal and $n$ is the number of blob size labels we use (in the present analysis we use $n=5$, therefor we have 5 different blob amplitudes in the analysis).
\begin{figure}
    \includegraphics[width=83mm]{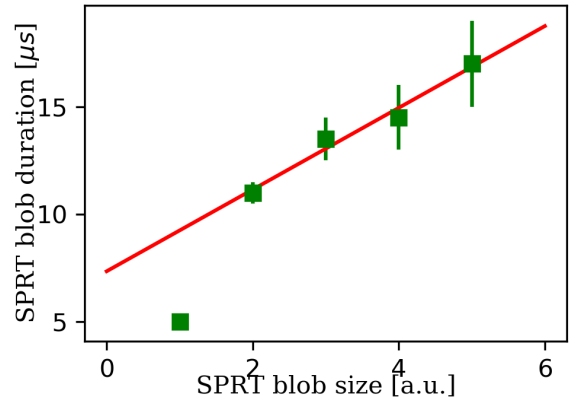}
    \caption{The observed linear correlation between the blob amplitude and the blob duration using the gSPRT method.}
    \label{fig:SPRT_blob_detection}
\end{figure}
There has been observed a strong correlation between the duration time of blobs as measured by the gSPRT method and the blob intensity as shown in Figure \ref{fig:SPRT_blob_detection}. We show the outlier point at the 'size 1' blobs as a warning sign which has to be taken into account when low amplitude blobs are considered. As the blobs (filaments) are considered to be 'coherent structures', the blob amplitudes, with good approximation are constant as they travel radially across the detection channel, therefore the measured blob duration is determined by the blob velocity. The analysis presented in Figure \ref{fig:SPRT_blob_detection} shows that with increasing blob amplitude the blob duration increases therefore the velocity decreases. Nevertheless we should note that the above statement does not necessarily imply a relation between the blob velocity and the spatial ($\mathbf{B}_\perp$) blob size. 

\subsubsection{The threshold method and conditional averaging.}\label{sec:conditonal_average}

\begin{figure}
    \includegraphics[width=83mm]{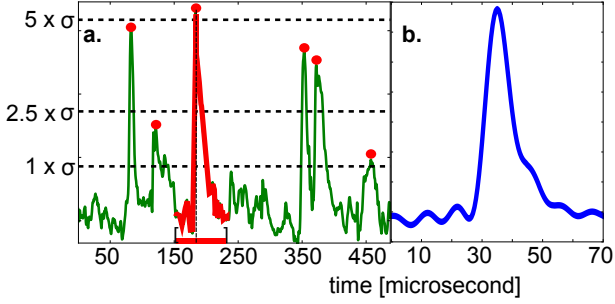}
    \caption{(a) Blob detection using the threshold method and time window selection for conditional average. (b) Conditional average blob, averaged over 420 events ($3\cdot 10^3~\mathrm{blob/sec}$) with amplitude  $>2.5\times\sigma$.}
    \label{fig:ca_explained}
\end{figure}

As it has been mentioned above, the threshold-based event detection followed by conditional averaging (CA) is the conventional approach in the field of scrape-off layer fluctuation studies. In this section we  briefly summarize our algorithm before we dive into data analysis. The algorithm starts with setting a reasonable threshold level, usually given in terms of standard deviation (e.g. $2.5\times\sigma$, see Figure \ref{fig:ca_explained}), then the signal is being scanned starting from the beginning until a peak is found above the threshold. Around the peak the signal is cut out within a $\Delta t$ time window (in the case seen on Figure \ref{fig:ca_explained}, $\Delta t=70~\mu s$) and saved for further analysis.

This procedure is repeated for all peaks above the threshold. As a result a collection of $\Delta t$ long time series is created. Conditionally averaged blob is obtained after averaging over this  ensemble of time series. 

\begin{figure}
\centering
    \includegraphics[width=83mm]{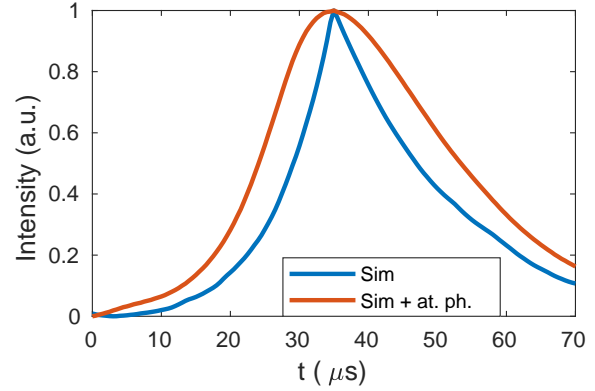}
\caption{Conditionally averaged waveforms calculated from simulated density signal (blue line) and after adding the atomic physics effects (red line).}
\label{fig:sim_ca}
\end{figure}

In Figure \ref{fig:sim_ca} we show the result of the conditional averaging procedure applied to simulated density time traces. The comparison of the "density blob" and the "light emission blob" is plotted in terms of the conditional average waveforms.  
The smearing effect of the plasma-beam interaction can be observed in the widening of the waveform. The rising exponent of the fitted exponential is increased by 72\% while the falling by 37\%. 

\subsection{Experimental observation of SOL fluctuations}\label{sec:experiment}

In this section we present the results of basic statistical analysis of intermittent SOL fluctuations (blobs and holes) as measured by the Li-BES system at COMPASS.     

\begin{figure}
\centering
    \includegraphics[width=83mm]{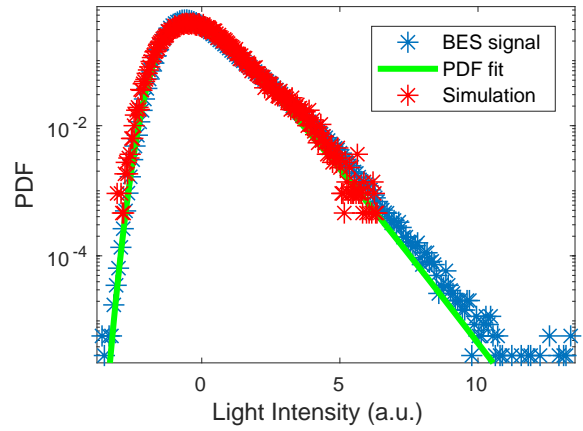}
\caption{Comparison of the measured PDFs in the SOL with the simulated synthetic signal and the result of the curve fitting procedure.}
\label{fig:fitting_pdfs}
\end{figure}

After filtering the raw Li-BES data to the relevant $1-100~\mathrm{kHz}$ frequency range, we calculated the amplitude histogram (PDF) of the Li-light fluctuations. It can be seen in \autoref{fig:fitting_pdfs} that the PDF is clearly positively skewed suggesting the presence of large amplitude events (filaments, blobs) in our experimental data. It is also clear that the experimental PDF does not fit the expected theoretical Gamma distribution described in \cite{garcia2012}. Nevertheless it was possible to develop a fitting function as a combination of two two-parameter distributions: a Gamma function and a Gaussian function which gave excellent fit to the data. It was also possible to deduce, directly from the experimental data, the input parameters for  synthetic signal simulations, namely the waiting time distribution and the average blob duration. Running the simulation with such parameters, the resulting PDF is in very good agreement with the actual measurement (see the red points in \autoref{fig:fitting_pdfs}), suggesting that our simulation approach is reliable. We have to note that the experimentally observed intermittency is lower than the Garcia-limit, yet the final conclusion of the model, that the signal PDF fits a Gamma distribution, still holds.  

As we have already showed in \autoref{sec:blob_simulation}, in the simulation \emph{hole artifacts} can appear behind the maximum of the Li-light emission profile. This will present a challenge to experimental data analysis due to the fact that the real holes (density deficit) are expected around the separatrix, where the blobs are created. 

\begin{figure}
\centering
    \includegraphics[width=83mm]{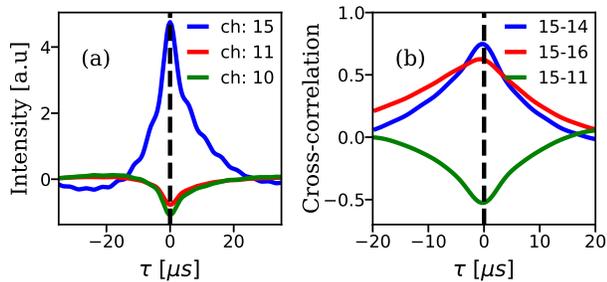}
\caption{(a) Conditional average waveforms measured at different radial positions, reference being channel \# 15. (b) Measured cross-correlation functions with reference channel \# 15. }
\label{fig:fake_holes}
\end{figure}

Indeed, hole artifacts can be found in the experimental data as well, as it can be seen in \autoref{fig:fake_holes}, where we show the conditional average  waveforms for different radial channels (\autoref{fig:fake_holes}.a). Here the reference channel \# 15 is located in the scrape-off layer while channels \# 10 and \# 11 are behind the light profile maximum (channel \#12). There is no time delay observed between conditionally averaged waveform maxima, showing that these events appear at the same time in the signal. The same conclusion can be drawn from the cross-correlation functions shown in \autoref{fig:fake_holes}.b. It is worth mentioning that reference channel is also \# 15 and time shifts relative to zero time lag can be observed for the SOL channels (\# 14, \#16) indicating radial propagation of the blobs, but there is no shift for channel \# 11, which is consistent with the presence of hole artifacts.  

\begin{figure*}
\centering
    \includegraphics[width=166mm]{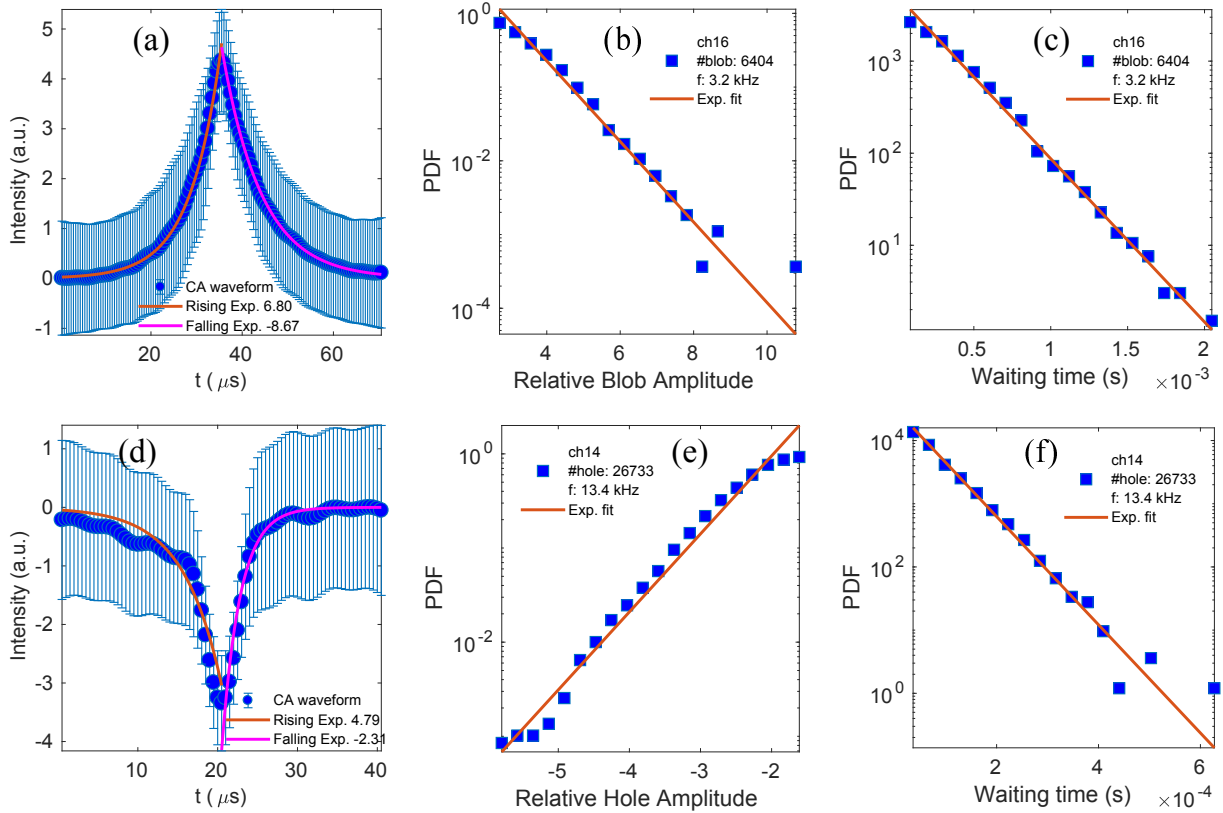}
\caption{Fundamental statistical properties of  blobs and holes measured by COMPASS Li-BES.}
\label{fig:blob_holes_statistics}
\end{figure*}

In order to meet the challenge presented in the previous paragraph,  the statistical analysis of the blob and hole events should start with a procedure sorting out the artifacts. This has been done based on the following process: if the algorithm finds a hole in a given channel and simultaneously  finds a blob somewhere else it qualifies the hole as a fake. Fake blob appearance is also possible, if for some reason the density drops locally in the SOL, it can cause a simultaneous density increase behind the emission profile. As the analyzed discharges are made to be as much identical as possible, we are able to collect large amount of events (few thousand). In \autoref{fig:blob_holes_statistics} we summarize the essential results about the statistical properties of these fluctuation events. In \autoref{fig:blob_holes_statistics}.a-b the conditional average waveforms are plotted for blobs in the  SOL and holes around the separatrix. The average blob duration is $\le 20~\mu s$ while the hole duration time is $\le 10~\mu s$. We have to note though that these values contain contributions from  different size blobs, and as it was shown on \autoref{fig:SPRT_blob_detection} the higher amplitude blobs have longer duration times, which is the reason for the relatively large errorbars. Fitting double exponential to the conditional average waveforms, the rising and falling times can be obtained. As a general statement we can say that the blob/hole waveform rises approximately two times faster than it decays. The \autoref{fig:blob_holes_statistics}.c-d  shows the amplitude distribution  for large events. According to Garcia model the amplitude distribution can be described by an exponential function, which indeed is the case for our Li-BES measurements. The above mentioned stochastic modeling predicts exponential distribution for the waiting times as well. Both the blobs and the holes detected in our experimental data have waiting times following exponential distribution as seen in \autoref{fig:blob_holes_statistics}.e-f. From the waiting time distribution the blob/hole detection rate can be inferred, this gives for blobs, $\lambda_\mathrm{blob}\approx 3\cdot10^3~\mathrm{s}^{-1}$ and for holes,  $\lambda_\mathrm{hole}\approx13\cdot10^3~\mathrm{s}^{-1}$.

\section{Summary}
We have demonstrated that the COMPASS Li-BES system can be used as a non-perturbative means for plasma fluctuation study, although the interpretation of the data is affected by the plasma-beam interaction. 

Our sensitivity study shows that for given density profile shape with density perturbation applied at a given radial position, the amplitude response is proportional to the perturbation amplitude. On the other hand if we keep the perturbation amplitude constant, the light response is more sensitive to outer perturbations and for density profiles with larger gradients. We have also found that the localization (maximum shift and broadening) of the response is better for outer channel perturbations and it practically does not depend on saturation density.

We constructed a synthetic diagnostic based on two-dimensional numerical simulation of moving random structures with prescribed spatial and temporal shapes. From the simulated data, using a collisional-radiative model we have generated synthetic Li-BES signals. The analysis revealed the appearance of artificial events (holes and blobs) due to the plasma-beam interactions. These "fake" events should not be included in the analysis. We have shown that the PDFs, therefore the statistics, are not affected by the atomic physics effects, but the photon noise and electrical noises change the statistics. 

We have shown that beside the generally used threshold method, the sequential probability ratio test method can be successfully applied for blob detection, having the advantage of more accurate determination of event duration directly from the signal statistics. 

We have also demonstrated that the Li-BES data of intermittent fluctuations in the scrape-off layer can be described by the stochastic Garcia model for convoluted PDFs assuming finite intermittency. It is also interesting to note that we have found a change in the correlation-functions of Li-light fluctuations as we approach the separatrix from the scrape-off layer. The origin of the wave-like correlations at the last closed flux surface is not yet clear. Quasi-two dimensional Li-BES measurements are planned for further investigation of turbulence phenomena at COMPASS.

\section*{Acknowledgement}

This work was partly supported by Czech Science Foundation Project No. GA16-25074S.\\
This work was also partialy supported by project No. CZ.02.1.01/0.0/0.0/16\_019/0000768 and MEYS project LM2015045.
\bibliography{ref.bib}
\bibliographystyle{IEEE}

\end{document}